\documentclass[aps,notitlepage]{revtex4}
\usepackage{graphicx}
\usepackage{subfigure}
\begin{document}
\title{Classification of materials with divergent magnetic Gr\"uneisen parameter}\author{Philipp Gegenwart}
\address{EP VI, Center for Electronic Correlations and Magnetism, Institute of Physics, Augsburg University, 86159 Augsburg, Germany}
\vspace{10pt}

\begin{abstract}
At any quantum critical point (QCP) with a critical magnetic field $H_c$, the magnetic Gr\"uneisen parameter $\Gamma_{\rm H}$, which equals the adiabatic magnetocaloric effect, is predicted to show characteristic signatures such as a divergence, sign change and $T/(H-H_c)^\epsilon$ scaling. We categorize thirteen materials, ranging from heavy fermion metals to frustrated magnets, where such experimental signatures have been found. Remarkably, seven stoichiometric materials at ambient pressure show $H_c=0$. However, additional thermodynamic and magnetic experiments suggest that most of them do not show a zero-field QCP. While the existence of a pressure insensitive ``strange metal'' state is one possibility, for some of the materials $\Gamma_{\rm H}$ seems influenced by impurities or a fraction of moments which are not participating in a frozen state. To unambiguously prove zero-field and pressure sensitive quantum criticality, a $\Gamma_{\rm H}$ divergence is insufficient and also the Gr\"uneisen ratio of thermal expansion to specific heat must diverge.
\end{abstract}

\maketitle

\section{Introduction}

Motivated by the theoretical proposal of a universal divergence in the approach of a quantum critical point (QCP)~\cite{zhu,Garst05}, the Gr\"uneisen ratio as well as its counterpart, the magnetic Gr\"uneisen parameter, have been studied for various strongly correlated electron systems in recent years~\cite{Kuechler03,Kuechler04,Kuechler_Physica06,Kuechler07,Gegenwart10,Gegenwart16}. The Gr\"uneisen ratio
\begin{equation}
\Gamma\sim{\alpha\over{C}}{\sim {(dS/dp)\over T(dS/dT)}}
\end{equation}
of thermal expansion $\alpha$ divided by specific heat $C$ is temperature independent, if the entropy $S$ can be scaled in the form $f(T/E^\star)$ with single energy scale $E^\star$, for example $T_c$ for a classical phase transition or the Fermi temperature for a metal~\cite{zhu}. In case of a QCP there is however no fixed energy scale and as detailed below the Gr\"uneisen ratio is expected to display a universal divergence.

While the Gr\"uneisen ratio probes the sensitivity to changes of pressure, in many cases the magnetic field has been used as control parameter. The magnetic Gr\"uneisen parameter is defined as
\begin{equation}
\Gamma_{\rm H}=-{{dM/dT}\over{ C}}={1\over T}({dT \over dH})_S,
\end{equation}
where $M$, $T$, $C$ and $S$ denote the magnetization, temperature, heat capacity and entropy, respectively. Whenever a QCP is driven by a magnetic field $H$ the control parameter $r$ can be defined, as $r=(H-H_c)/H_0$, where $H_c$ is the critical field and $H_0$ denotes a constant (which is a characteristic field of the material)~\cite{Garst05}. In the scaling regime close to a QCP, the correlation length of order parameter fluctuations $\xi\sim|r|^\nu$ diverges with exponent $\nu$. The temporal criticality is characterized by $\xi_\tau\sim \xi^z$ with dynamical critical exponent $z$. The Gr\"uneisen parameter diverges upon approaching the QCP at $r=0$ following $T^{-1/(\nu z)}$ or respectively at $T=0$ according to $r^{-1}$ with universal prefactor $G_r=\nu(d-z)$~\cite{zhu,Garst05}.

\begin{table}[h!]
\centering
\caption{List of materials for which a divergence of the magnetic Gr\"uneisen parameter and/or temperature over magnetic field scaling in thermodynamic properties has been observed. While the first four listed materials display scaling with a finite critical field for the remaining ones $H_c=0$ within experimental accuracy. The critical field $H_c$ and the prefactor $G_r$ are obtained from $\Gamma_{\rm H}=-G_r/(H-H_c)$. The scaling exponent $\epsilon$ is obtained from $T/(H-H_c)^\epsilon$ scaling. \\}
\label{tab:table1}
\begin{tabular}{c|c|c|c|c|c}
\shortstack[c]{Material\\~\\~} & \shortstack[c]{$x$ in\\ $\Gamma_{\rm H}\sim T^{-x}$\\~} & \shortstack[c]{$\mu_0 H_c$ (T)\\~} & \shortstack[c]{$G_r$ \\~} & \shortstack[c]{scaling\\ exponent $\epsilon$ \\~}& \shortstack[c]{Ref.\\~} \\
\hline \hline
YbRh$_2$Si$_2$ & 2 & 0.065 & $-0.3$ & 1 & \cite{Tokiwa09,Custers}\\
YbAgGe & 1.8 & 4.8 & $-0.31$ & 1.1 & \cite{TokiwaYbAgGe}\\
Sr$_3$Ru$_2$O$_7$ & 2.66 & 7.53 and 7.845 & $-0.17$ & 0.75 & \cite{TokiwaSr327}\\
Li$_2$Ir$_{0.45}$Ti$_{0.55}$O$_3$ & 1.7 & 0.2 & $-0.6$ & 0.86 & \cite{Manni}\\ 
\hline
Ce(Ni$_{0.935}$Pd$_{0.065}$)$_2$Ge$_2$ & 2.4 & 0 & & & \cite{Gegenwart10}\\
Yb$_{0.81}$Sc$_{0.19}$Co$_2$Zn$_{20}$ & 1.0 & 0 & $-0.52$ & & \cite{Tokiwa16}\\ 
CeCoIn$_5$ & 1.33 & 0 & $-0.85$ & 1.5 & \cite{TokiwaCeCoIn5}\\
CeRhSn & 2.7 & 0 & & & \cite{TokiwaCeRhSn}\\
$\beta$-YbAlB$_4$ & 2 &0 & & 1 & \cite{Matsumoto}\\
Au-Al-Yb quasicrystal & 1.8 &0 & $-0.3$ & 1 & \cite{Deguchi}\\
YbCo$_2$Ge$_4$ & 2.5 & 0 & $-0.095$ & 1 & \cite{Sakai}\\
Pr$_2$Ir$_2$O$_7$ & 1.5 & 0 & $-0.25$ & 1.33 & \cite{Tokiwa14}\\
Na$_4$Ir$_3$O$_8$ & 3 & 0 & $-0.25$ & 0.66 & \cite{Singh}\\
\end{tabular}
\end{table}

Such divergences have been found in magnetic Gr\"uneisen parameter measurements on various strongly correlated electron systems in recent years, which are listed in Table~1~\cite{Tokiwa09,TokiwaYbAgGe,TokiwaSr327,Manni,Tokiwa16,Gegenwart10,TokiwaCeCoIn5,TokiwaCeRhSn,Matsumoto,Deguchi,Sakai,Tokiwa14,Singh}. 
The table does not include low dimensional spin systems for which the magnetic Gr\"uneisen parameter has been investigated experimentally and theoretically near field-induced quantum phase transitions~\cite{Wolf,Ryll,Galisova}. Due to perturbing interactions it is often difficult to approach the asymptotic low-temperature power law behavior of $\Gamma_{\rm H}$ for such systems. The first four entries of the table concern materials with a QCP at finite magnetic field, while in all subsequent cases the critical field is zero. The paramagnetic heavy fermion metals CeNi$_2$Ge$_2$ and YbCo$_2$Zn$_{20}$ have been driven towards a zero-field QCP by suitable chemical substitution~\cite{Gegenwart10,Tokiwa16}. However, there is a remarkably huge number of undoped materials, for which the critical field is zero within experimental accuracy. It seems very surprising, that a compound is located accidentally at a QCP without need to fine tune composition, pressure or magnetic field. As we will discuss in this article, indeed many of those materials do not display generic (pressure-sensitive) quantum criticality and the magnetic Gr\"uneisen parameter divergence has other origin. Aim of this article is to categorize materials for which a divergence of the magnetic Gr\"uneisen parameter and temperature over field scaling with zero critical field has been reported. In the following, we first discuss field-induced quantum criticality (section 2) and quantum criticality in geometrically frustrated systems (section 3). The subsequent section 4 is dedicated to quantum criticality and superconductivity. Section 5 deals with materials that do not show pressure-sensitive quantum criticality but likely some pressure-insensitive strange metal states. In section 6, divergent behavior of the magnetic Gr\"uneisen parameter in disordered frustrated magnets, which display spin-glassy effects is discussed before we draw a conclusion in section 7.  

\section{Finite-induced quantum criticality}

YbRh$_2$Si$_2$ is a prototype heavy fermion metal, which displays a field-induced QCP, related to the suppression of very weak antiferromagnetic ordering below $T_N=70$~mK~\cite{Tokiwa09,Custers}. For fields $H>H_c$ the magnetic Gr\"uneisen parameter displays a $-G_r/(H-H_c)$ dependence and the value of the dimensionless constant $G_r=-0.3$ equals the exponent of the quasiparticle mass divergence as predicted by theory~\cite{zhu,Garst05}. However, $\Gamma_H(T)$ at $H=H_c$ has shown a crossover near 0.3~K which may be related to competing antiferro- and ferromagnetic fluctuations and challenges the interpretation of the results within the quantum critical regime~\cite{Tokiwa09}.

Field-induced quantum criticality can also result from the suppression of the critical temperature of a first-order metamagnetic transition towards absolute zero temperature~\cite{Millis02}. This has been proposed for the bilayer strontium ruthenate Sr$_3$Ru$_2$O$_7$~\cite{Grigera01}. Very recently a careful study of the magnetic Gr\"uneisen parameter has revealed that this material actually realizes a more complicated scenario, with two subsequent field-induced QCPs (for the critical fields see Table 1)~\cite{TokiwaSr327}. The $\Gamma_H$ data also allowed to determine
scaling regimes associated with both QCPs and the phase space where scaling fails due to the interference of both instabilities. The observed scaling exponents are consistent with the itinerant theory for metamagnetic ($z=3$) quantum criticality in $d=2$~\cite{Millis02,GegenwartSr327}.

For more details on Gr\"uneisen studies on field- and concentration tuned quantum criticality in various heavy fermion metals, including graphs for YbRh$_2$Si$_2$, YbAgGe and Sr$_3$Ru$_2$O$_7$, we refer to the recent review~\cite{Gegenwart16}.

\section{Quantum criticality and geometrical frustration}

\begin{figure}
\begin{center}
\includegraphics[width=\textwidth]{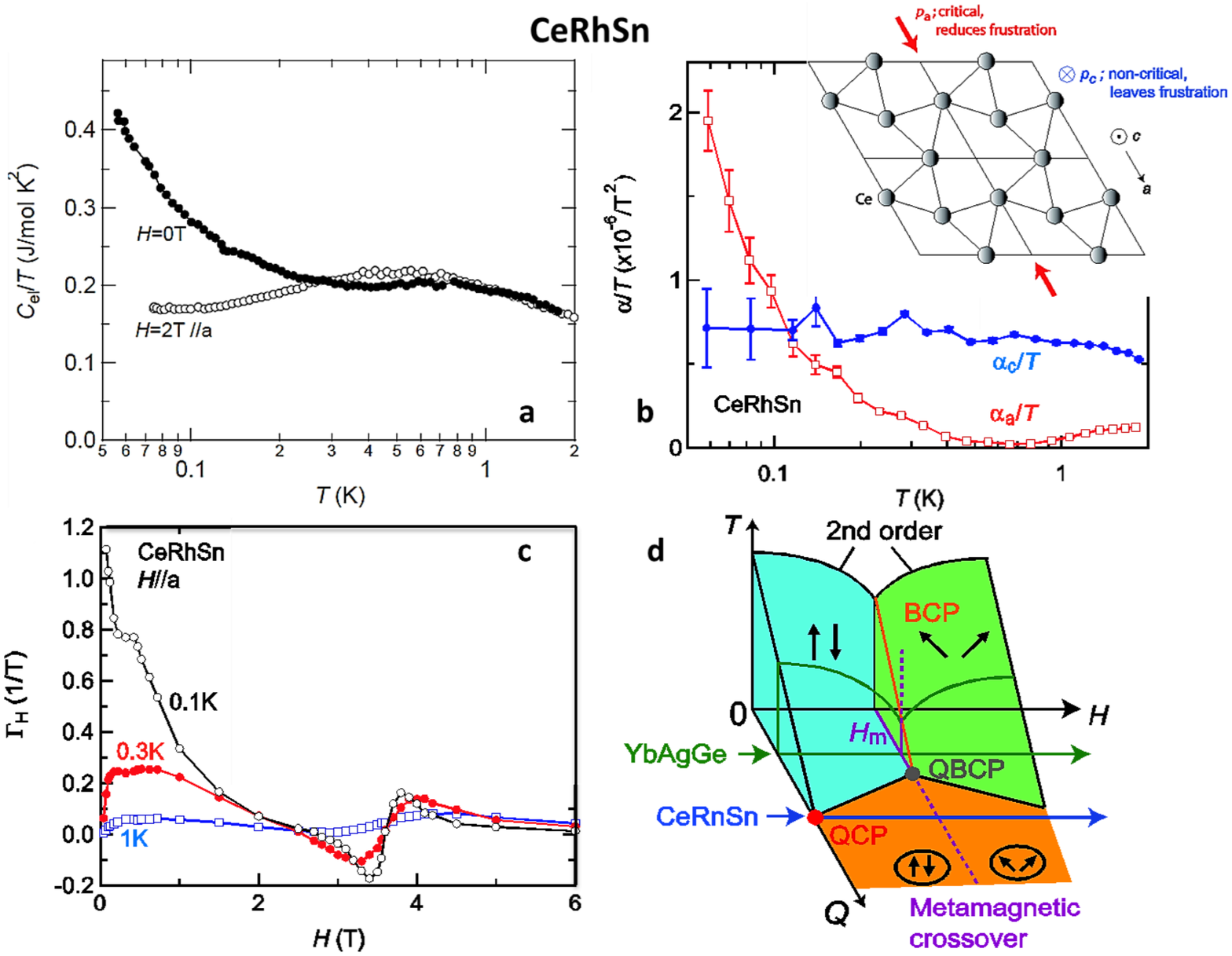}
\caption{Evidence for a frustration induced QCP in hexagonal CeRhSn with distorted Kagome configuration of Ce atoms (cf. right inset)~\cite{TokiwaCeRhSn}. Specific heat coefficient (a) and linear thermal expansion coefficient (b) vs. $T$ (on log scale). Magnetic Gr\"uneisen parameter vs field (c). Schematic 3D temperature vs. magnetic field vs. frustration phase diagram, where $Q$ indicates the strength of quantum fluctuations induced by geometrical frustration (d).} \label{CeRhSn}
\end{center}
\end{figure}

For Kondo lattice metals, a ``global phase diagram'' with two parameters has been proposed, which are (i) the relative strength of the Kondo compared to the intersite interaction and (ii) the strength of quantum fluctuations~\cite{Si}. Enhanced quantum fluctuations, caused by low-dimensional or frustrated interactions, are counter-acting the Kondo singlet formation and should ultimately lead to a spin liquid state with local magnetic moments~\cite{Vojta}. Therefore a novel QCP arising from strong geometrical frustration has been expected. The effect of geometrical frustration has been studied recently in Kondo lattices crystallizing in the ZrNiAl structure, where the f-moments are residing on a distorted Kagome lattice stacked along the c-axis. For CePdAl evidence for partial magnetic ordering has been found~\cite{Donni}, arising from frustration. By chemical substitution this system can be tuned towards a QCP~\cite{Fritsch}, whose nature has most recently been investigated by the magnetic Gr\"uneisen parameter. In addition to the generic signatures of field-induced quantum criticality, a concentration independent broadened peak in $\Gamma_H(H)$ has been found and ascribed to correlations of the frustrated moments~\cite{Fritsch16}. YbAgGe is an isostructural Yb-based Kondo lattice, which features a couple of almost degenerate magnetic phases in the temperature field phase diagram~\cite{Budko}. Of particular interest is a first-order transition between two magnetically ordered phases near 4.8~T. Measurements of the magnetic Gr\"uneisen parameter provide evidence for quantum bicritical behavior associated with this transition~\cite{TokiwaYbAgGe}. As sketched in Fig. 1(d), a scenario has been proposed, where such a field-induced QCP arises from the suppression of a line of first-order spin-flop transitions towards absolute zero temperature. Further enhancing the strength of quantum fluctuations would then result in a complete suppression of long-range order at zero field and metamagnetic crossover at the field of the original first order transition.

The isostructural paramagnetic heavy-fermion metal CeRhSn~\cite{Kim} displays such behavior. At $H=0$ and below 1~K the heat capacity coefficient, the Gr\"uneisen ratio, as well as, the magnetic Gr\"uneisen parameter all diverge, evidencing a zero-field QCP~\cite{TokiwaCeRhSn}. Furthermore, linear thermal expansion is highly anisotropic and displays quantum criticality only perpendicular to the $c$-axis. Since linear thermal expansion is given by the uniaxial pressure derivative of the entropy and since entropy for quantum critical materials has critical and non-critical contributions~\cite{zhu}, the data indicate, that quantum criticality only couples to in-plane uniaxial pressure but not to $c$-axis uniaxial pressure. This is in accordance with quantum criticality driven by geometrical frustration, since the latter is not affected by $c$-axis uniaxial pressure~\cite{TokiwaCeRhSn}.

\begin{figure}
\begin{center}
\includegraphics[width=\textwidth]{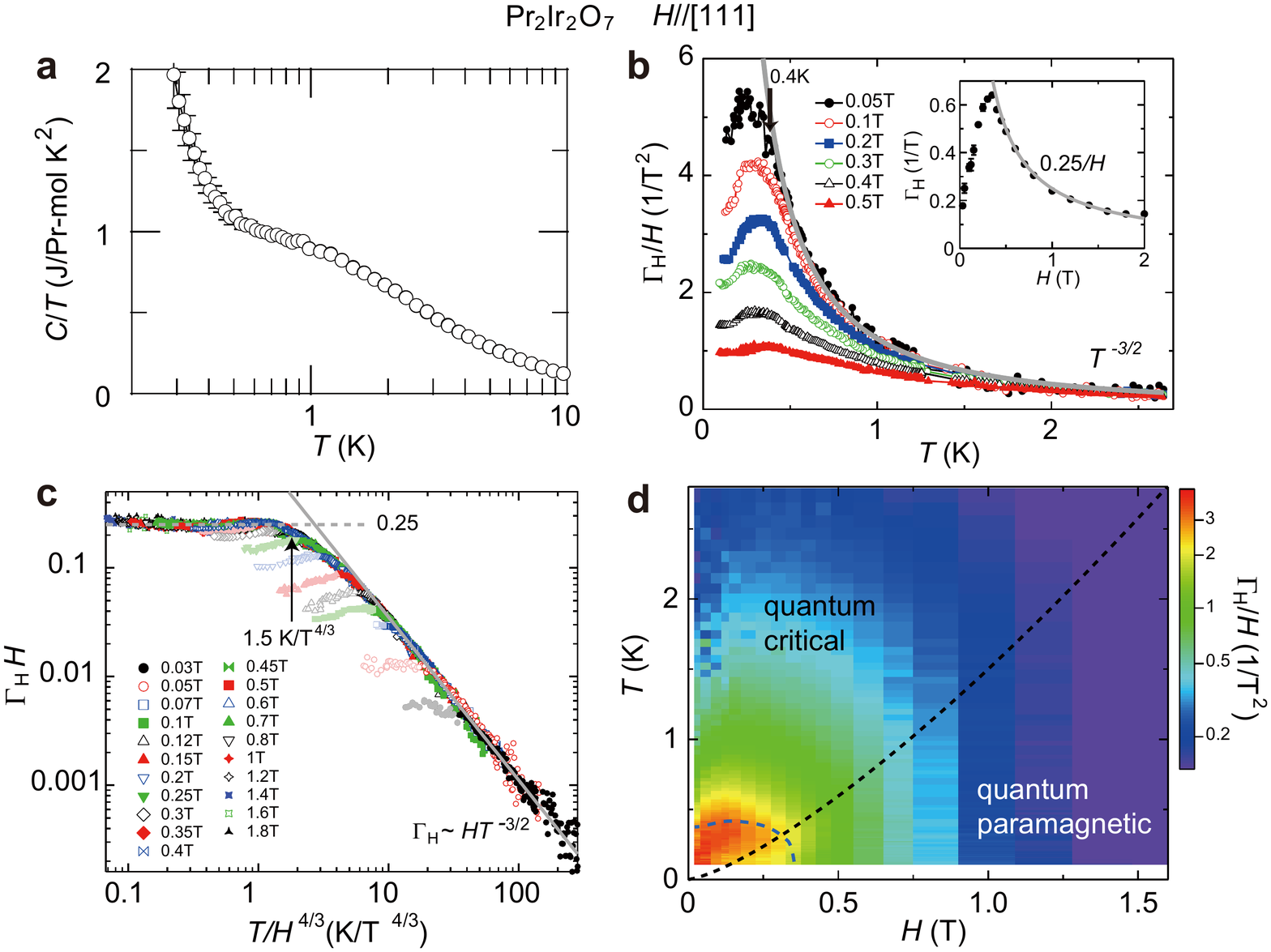}
\caption{Zero-field quantum criticality in pyrochlore Pr$_2$Ir$_2$O$_7$~\cite{Tokiwa14}. Specific heat (after subtraction of nuclear, phonon and crystal field contributions) as $C/T$ vs. $T$ (on log scale) (a) and magnetic Gr\"uneisen parameter as $\Gamma_{\rm H}/H$ vs. $T$ for various different magnetic fields, applied along the [100] direction (inset displays the field dependence of $\Gamma_{\rm H}$ in the limit of $T\rightarrow 0$) (b). Scaling behavior of $\Gamma_{\rm H}$ (c) and color coded contour plot of $\Gamma_{\rm H}$ in the temperature-field plane (d). The dotted line in (d) marks the crossover between the quantum critical and quantum paramagnetic regime, as determined from the saturation of $\Gamma_{\rm H}(T)$, indicated by the black arrow in (c). The blue dotted line in (d) encloses the regime where $\Gamma_{\rm H}$ deviates from scaling, cf. data points in pastel color in (c).}
\end{center}
\end{figure}

Zero-field quantum criticality possibly related to strong geometrical frustration has also been found in the low-carrier density pyrochlore iridate Pr$_2$Ir$_2$O$_7$~\cite{Tokiwa14}. It has local Pr$^{3+}$ magnetic moments in tetrahedral spin-ice configuration with a small concentration of Ir 5d conduction electrons~\cite{Nakatsuji}. In contrast to classical spin ice, the non-Kramers doublet ground state is an effective 1/2 spin and quantum fluctuations may lead to a melting of the spin ice ground state at low temperatures. Respectively, the heat capacity coefficient does not decay exponentially as for classical spin ice, but increases below 1~K~\cite{Tokiwa14} (cf. Fig. 2(a)). As shown in Fig. 2(b), the magnetic Gr\"uneisen parameter displays a divergence towards zero magnetic field down to temperatures of about 0.4~K, below which deviation is found. This is accompanied by characteristic temperature over magnetic field scaling (panel c)) which describes the crossover from the quantum critical into the quantum paramagnetic regimes of the phase diagram (cf. panel (d)) except for temperatures below 0.4~K and fields below 0.4~T. Thus, zero-field quantum criticality is avoided by the formation of some yet unknown state at lowest $T$ and $H$.

\section{Quantum criticality and superconductivity}

\begin{figure}
\begin{center}
\includegraphics[width=\textwidth]{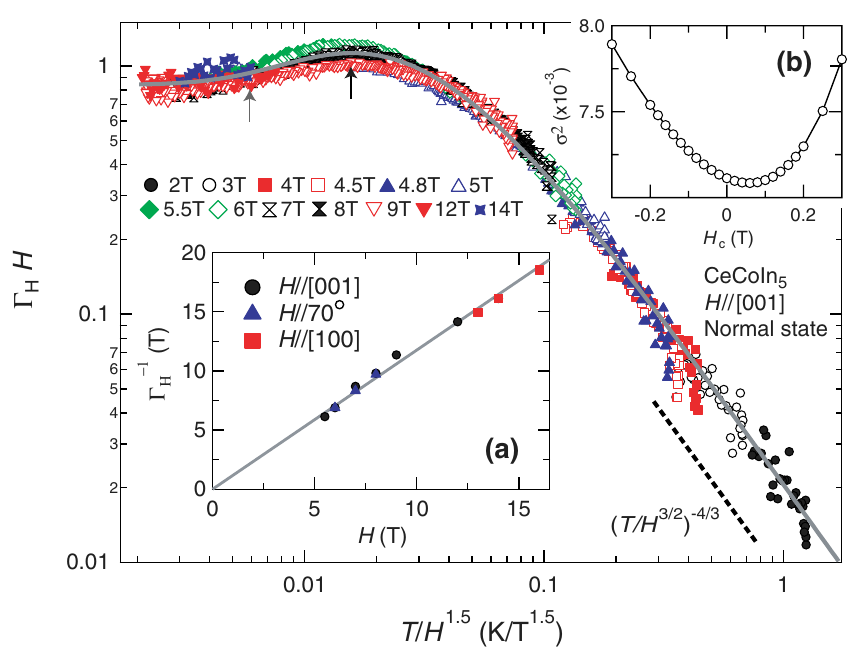}
\caption{Temperature over field scaling of the magnetic Gr\"uneisen parameter of CeCoIn$_5$ as $\Gamma_{\rm H}H$ vs. $T/H^{1.5}$. Inset (a) displays $\Gamma_{\rm H}^{-1}$ vs. $H$ for data obtained in various field orientations and the line indicates $\Gamma_{\rm H}=0.85/H$. The weighted mean-square deviation of the data from a phenomenological function (cf. grey line in the main plot, with maximum and inflection points indicated by black and grey arrows, respectively) is shown in the inset (b) as function of the critical field~\cite{TokiwaCeCoIn5}.} \label{CeRhSn}
\end{center}
\end{figure}

Unconventional superconductivity often occurs in the vicinity of competing magnetic states suggesting a relation to quantum criticality. Here, we focus on measurements of the magnetic Gr\"uneisen parameter in the heavy fermion superconductor CeCoIn$_5$ with $T_c=2.3$~K and upper critical field of 5 and 12~T for fields along and perpendicular to the tetragonal $c$-axis, respectively~\cite{Petrovic}. Various transport and thermodynamic properties have shown pronounced non-Fermi liquid behavior in the normal state of CeCoIn$_5$ and proposed a field-induced QCP~\cite{Bianchi,Paglione,SSingh,Donath,Zaum}. As displayed in the inset (a) of Fig. 3, the magnetic Gr\"uneisen parameter diverges as $\Gamma_{\rm H}\sim 1/H$ for different field orientations in the normal state~\cite{TokiwaCeCoIn5}. Extrapolation of this dependence to fields below the upper critical field and towards $H=0$ suggests a zero-field QCP. This is corroborated by the observation of temperature over field scaling, where the critical field is zero. For such a zero-field QCP, the scaling function for the free energy has no linear dependence on the control parameter (which is then just $H$), because the latter would yield a spontaneous magnetization~\cite{TokiwaCeCoIn5}. As a result of the (then leading) quadratic control parameter dependence, the scaling parameter $G_r=2\nu(d-z)$ now contains a factor 2 which is absent for the case of a finite-field QCP~\cite{TokiwaCeCoIn5}.

For the scaling, only data taken in the normal state could be included. The observed $1/H$ divergence of $\Gamma_{\rm H}$ for $H>5$~T is incompatible with previous claims of a critical field of 4 to 5~T~\cite{Bianchi,Paglione,SSingh,Donath,Zaum}. However, we note that the determination of $H_c=0$ relies on the linear extrapolation shown in the inset of Fig. 3 and therefore a small but finite critical field can hardly be excluded.

\section{Pressure-insensitive strange metal states}

\begin{figure}
\begin{center}
\includegraphics[width=\textwidth]{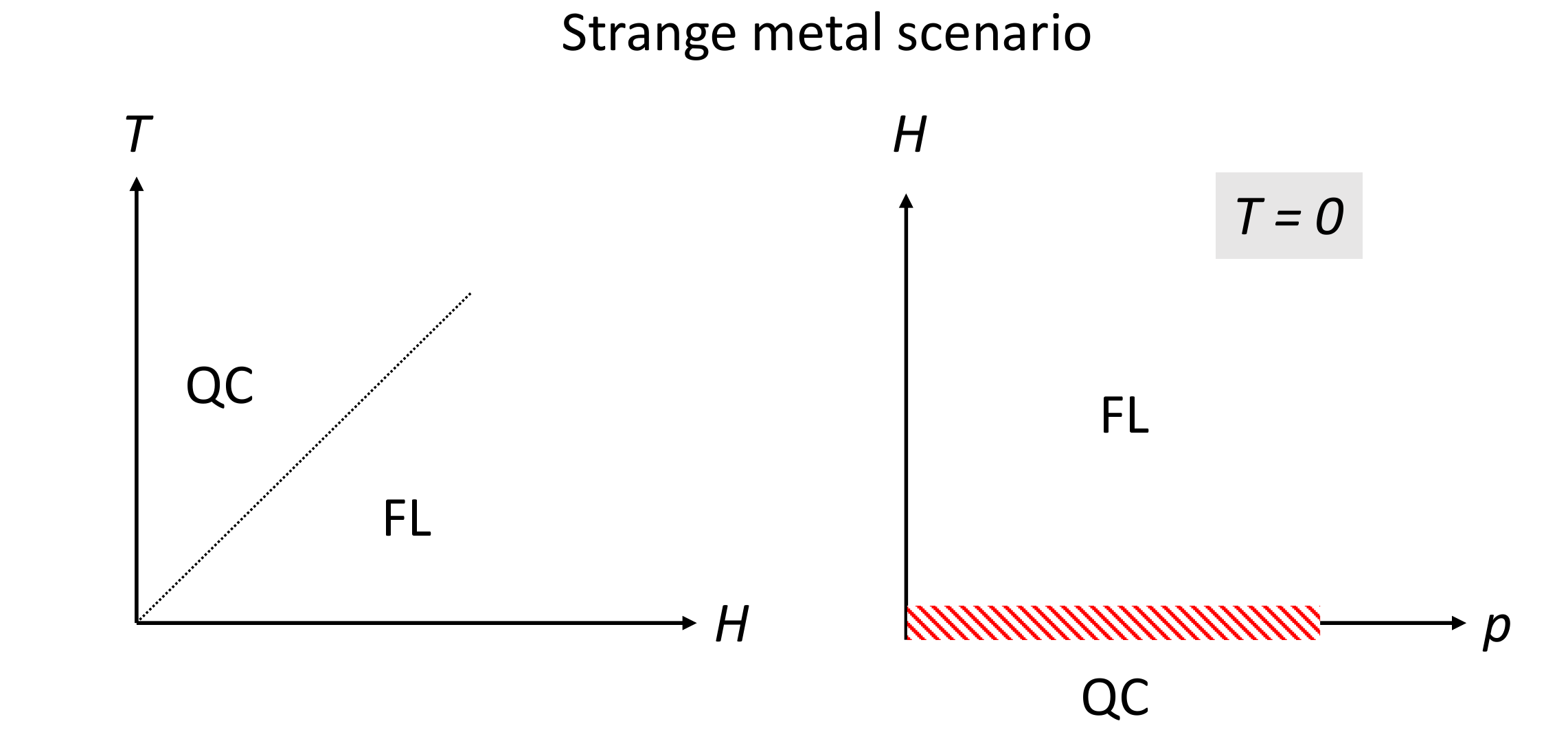}
\caption{Left: Temperature vs. magnetic field phase diagram with crossover between quantum critical (QC) and Fermi liquid (FL) regimes that terminas at the origin. Right: Magnetic field vs. pressure phase diagram at $T=0$ with extended QC phase at zero field.}
\end{center}
\end{figure}

We now discuss a couple of materials displaying signatures of zero-field quantum criticality that appears pressure insensitive as sketched in Figure 4. The main difference to a generic zero-field QCP is that there is no pressure dependence of the critical field, i.e., no ordered state is found in the immediate vicinity. Because of the insensitivity to pressure, the Gr\"uneisen ratio does not diverge in this case. As an example, the itinerant helical ferromagnet MnSi shows non-Fermi liquid behavior over an extended pressure range from 1.5 to at least 2.5~GPa with no indication of a Gr\"uneisen ratio divergence~\cite{Doiron,Pfleiderer}.

\begin{figure}
\begin{center}
\includegraphics[width=\textwidth]{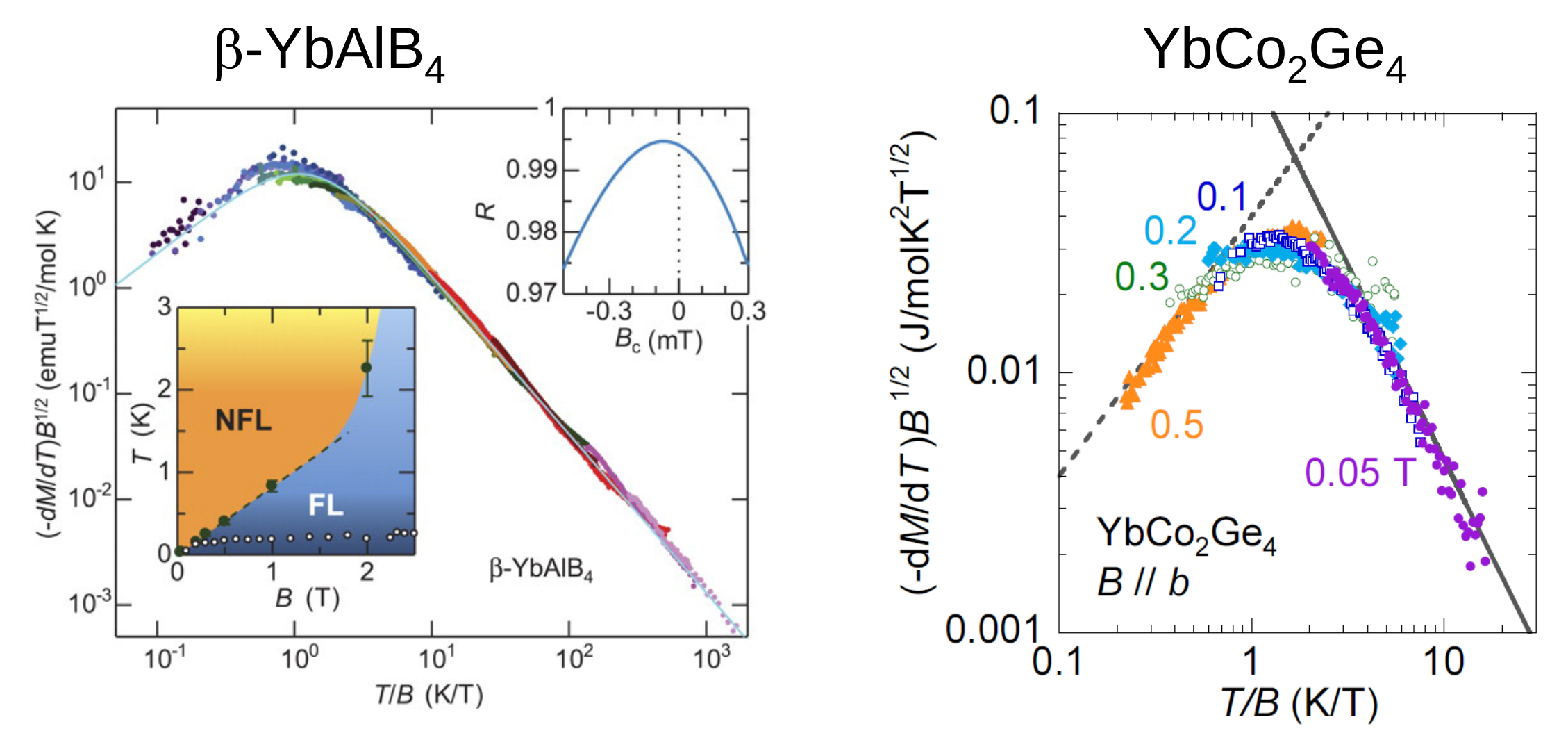}
\caption{Scaling behavior of the magnetization for $\beta$-YbAkB$_4$ (left panel)~\cite{Matsumoto} and YbCo$_2$Ge$_4$ (right panel)~\cite{Sakai}.}
\end{center}
\end{figure}

In $\beta$-YbAlB$_4$, as shown in Fig. 5 (left), thermodynamic properties at ambient pressure such as $dM/dT$ display $T/B$ scaling and indicate a zero field QCP~\cite{Matsumoto}. However, hydrostatic pressure experiments have revealed the onset of long-range ordering only at pressures beyond 2.5~GPa~\cite{Tomita}. The electrical resistivity indicates non-Fermi liquid behavior at zero pressure, which extends over a large pressure range. This behavior has been labelled as ``strange metal'' state. The non-Fermi liquid behavior is field sensitive but pressure independent as sketched in Fig. 4.

Au-Al-Yb quasicrystal displays a similar $T/H$ scaling and divergent magnetic susceptibility as $\beta$-YbAlB$_4$~\cite{Deguchi}. We have recently performed  thermal expansion measurements. The expansion coefficient $\alpha/T$ is less divergent than the heat capacity coefficient $C/T$. Thus there is no divergence in the Gr\"uneisen ratio, excluding a pressure sensitive QCP in accordance with the scenario of Fig. 4~\cite{quasicrystal}. 

Next we discuss recent results on tetragonal YbCo$_2$Ge$_4$. In this stoichiometric paramagnetic HF metal the electrical resistivity shows a non-Fermi $\Delta\rho\sim T^{1.4}$ dependence between 0.1 and 1.4~K hinting at NFL behavior, possibly related to quantum criticality~\cite{Kitagawa}. This is corroborated by a logarithmic divergence of the heat capacity coefficient and local spin susceptibility (from nuclear magnetic and quadrupole resonance) at elevated temperatures. We have recently performed a thermodynamic study down to 0.1~K~\cite{Sakai}. Interestingly, the magnetic Gr\"uneisen parameter diverges upon cooling at low fields, while the heat capacity coefficient $C/T$ saturates below 1~K. Although the strong divergence of $-dM/dT$ would suggest ferromagnetic quantum criticality, the latter can be excluded from the heat capacity data. Even more interestingly, $dM/dT$ shows exactly the same scaling as found previously for $\beta$-YbAlB$_4$, cf. Fig. 5~\cite{Sakai}. Furthermore, also this material shows long-range ordering only above 3~GPa. Therefore YbCo$_2$Ge$_4$ may also be described as ``strange metal''. Another possibility is that a small concentration of paramagnetic impurities which is hardly visible in heat capacity leads to the strong temperature dependence of $dM/dT$ and thus $\Gamma_H(T)$ down to the lowest measured temperature (but not asymptotic to $T=0$). In this scenario, the magnetic Gr\"uneisen parameter divergence would be extrinsic. However, the simplest description of independent paramagnetic impurities, i.e., a Curie law in the susceptibility and a $T^{-2}$ behavior in heat capacity, arising from the high-$T$ tail of a Schottky anomaly, yields a temperature independent magnetic Gr\"uneisen parameter. In any case the result shows, that divergent behavior of the magnetic Gr\"uneisen parameter alone is insufficient to claim quantum criticality. A thermodynamic proof of zero-field (and pressure sensitive) quantum criticality requires, that in addition to the magnetic Gr\"uneisen parameter also the Gr\"uneisen ratio of thermal expansion to specific heat diverges. For Au-Al-Yb such Gr\"uneisen ratio divergence is absent~\cite{quasicrystal}. For both $\beta$-YbAlB$_4$ and YbCo$_2$Ge$_4$ millikelvin measurements of the thermal expansion coefficient, using capacitive dilatometers, have not yet been possible, due to the lack of sufficiently large single crystals.

\section{Frustrated disordered magnets}

\begin{figure}
\begin{center}
\includegraphics[width=\textwidth]{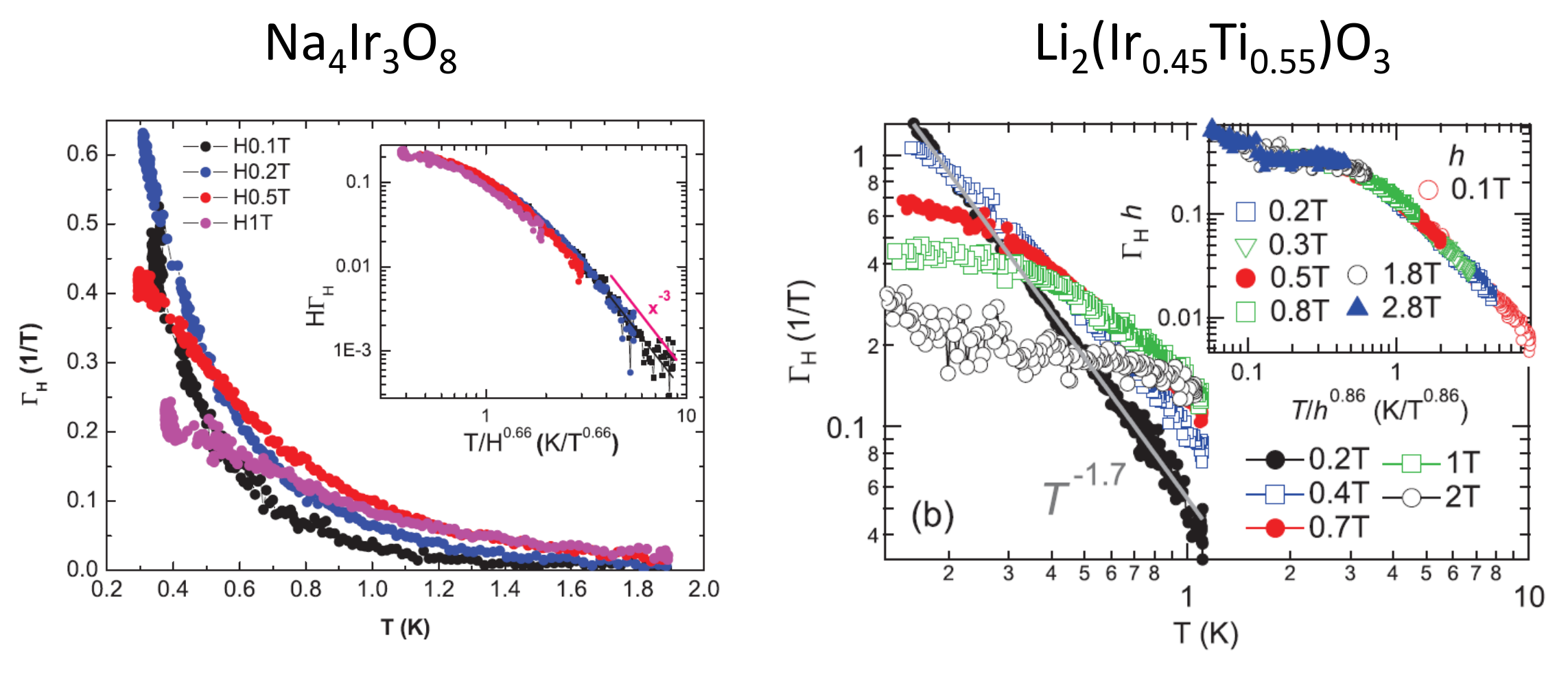}
\caption{Left: magnetic Gr\"uneisen parameter $\Gamma_{\rm H}$ vs. $T$ of Na$_4$Ir$_3$O$_8$. The inset displays temperature over magnetic field scaling, where the scaling parameter $x=T/H^{0.66}$~\cite{Singh}. Right: magnetic Gr\"uneisen parameter of Li$_2$(Ir$_{0.45}$Ti$_{0.55}$)O$_3$ with $T/h$ scaling (inset), where $h=H-$0.2~T.~\cite{Manni}}
\end{center}
\end{figure}

We now consider materials with divergent magnetic Gr\"uneisen parameter in which disorder plays an important role, evidenced by spin-glassy behavior. The two iridate materials~\cite{Manni,Singh} listed in table 1 belong to the class of frustrated spin-orbit Mott insulators. Hyperkagome Na$_4$Ir$_3$O$_8$ with effective 1/2 moments on a three-dimensional frustrated lattice has attracted considerable interest after the proposal of spin liquid behavior~\cite{Takagi}. It is a spinel oxide AB$_2$O$_4$ where the B-sublattice is occupied to 3/4 by Ir and 1/4 by Na atoms. Na$_4$Ir$_3$O$_8$ does not show long-range order down to low temperatures despite a large antiferromagnetic Curie Weiss temperature of about 650~K. However, the magnetic susceptibility, measured in a low field of 10~mT, has shown a small peak near 6~K, accompanied by hysteresis between warming and cooling~\cite{Takagi}. A detailed study of the $^{23}$Na and $^{17}$O nuclear magnetic resonance found evidence for disordered magnetic freezing of all Ir moments, which sets in below 7~K~\cite{Shockley}. The disorder, necessary for spin-glassy behavior, has been related to small local variations of the 3/4 Na occupancy of the B-sublattice, inducing a slight local charge disorder. A power-law behavior, found in the heat capacity below 4~K~\cite{Takagi,Singh} may be related to the free energy landscape typical for spin glasses~\cite{Shockley}. Interestingly, the magnetic Gr\"uneisen parameter $\Gamma_{\rm H}$, as shown in the left part of Fig.~6, displays a divergence upon cooling at low magnetic fields and temperature over magnetic field scaling~\cite{Singh}. However, in view of the spin-glass freezing below 7~K, intrinsic zero-field quantum criticality can clearly be excluded. Since the $\Gamma_H$ divergence occurs well below 7~K, it is likely related to a small concentration of magnetic moments, which are not yet frozen. Non-interacting paramagnetic impurities cannot cause a diverging $\Gamma_{\rm H}$ (see above). Possibly the observed divergence is caused by a small fraction of Ir moments which do not freeze below the spin-glass transition.

Similar results were also found in the partially diluted honeycomb iridate
Li$_2$(Ir$_{1-x}$Ti$_x$)O$_3$~\cite{Manni}. Unsubstituted $\alpha$-Li$_2$IrO$_3$ displays long-range ordering at 15~K. The latter becomes immediately suppressed by the partial dilution of the Ir site with non-magnetic Ti atoms and a spin-glassy ground state forms for $x\geq 0.09$. The freezing temperature decreases with increasing $x$ but persists to Ti concentrations up to 55\%. Thus, the material realizes a strongly smeared quantum phase transition. Temperature over magnetic field scaling and a divergence of the magnetic Gr\"uneisen parameter is shown on the right side of Fig. 6. A pressure sensitive QCP can be excluded from the observation of a temperature independent thermal Gr\"uneisen ratio~\cite{Manni}. Thus, the behavior of the magnetic Gr\"uneisen parameter is again likely related to a small fraction of Ir moments that remain unfrozen at low temperatures.

\section{Conclusion}

The Gr\"uneisen analysis provides a powerful thermodynamic tool to identify and characterize quantum criticality~\cite{zhu,Garst05,Gegenwart16}. It uses the (thermal) Gr\"uneisen ratio $\Gamma\sim \alpha/C$ and the magnetic Gr\"uneisen parameter $\Gamma_H=-(dM/dT)/C=T^{-1}(dT/dH)_S$ which can be determined by suitable techniques with high precision down to low temperatures. For field-induced QCPs, the signatures of the magnetic Gr\"uneisen parameter are (i) a divergence upon decreasing $T$ within the quantum critical regime, (ii) a sign change upon tuning the field across the critical field $H_c$ and (iii) $T/(H-H_c)^\epsilon$ scaling. Various materials for which such behavior was found are summarized in table 1. Among them seven stoichiometric materials have zero critical field. The determination of $H_c$ relies of course on the extrapolation to $T,H\rightarrow 0$, which in case of the superconducting CeCoIn$_5$ may be difficult. For the Kondo lattice CeRhSn~\cite{TokiwaCeRhSn} and the pyrochlore iridate Pr$_2$Ir$_2$O$_7$~\cite{Tokiwa14} zero-field quantum criticality seems to be related to the strong geometrical frustration.

In all other materials with zero critical field listed in table 1, a generic QCP at $H=0$ can be excluded. We have discussed a``strange metal'' scenario where zero-field non-Fermi liquid behavior appears insensitive of the application of pressure (cf. Fig. 4). This could be relevant for $\beta$-YbAlB$_4$, Au-Al-Yb quasicrystal and YbCo$_2$Ge$_4$. All of them show $T/H$ scaling with similar exponents, similar magnetic Gr\"uneisen parameter divergences and share the pressure insensitivity. The divergent behavior of $M(T)$ or magnetic susceptibility in these materials indicates strong ferromagnetic fluctuations. However, ferromagnetic quantum criticality cannot explain the observed behavior, e.g. the absence of a divergence in $C/T$ and $\Gamma(T)$ for YbCo$_2$Ge$_4$ and Au-Al-Yb quasicrystal, respectively. For YbCo$_2$Ge$_4$ the importance of paramagnetic impurities has been discussed~\cite{Sakai}, although non-interacting paramagnetic spins cannot give rise to a divergence of $\Gamma_H$.

A third class of materials displaying magnetic Gr\"uneisen parameter divergence and respective scaling but absence of generic quantum criticality consists of disordered highly frustrated magnets. These materials display spin-glassy freezing at low temperatures. The hyperkagome Na$_4$Ir$_3$O$_8$ is a three-dimensional geometrically frustrated magnet~\cite{Takagi} for which a frozen spin state below 7~K has been proven by recent NMR experiments~\cite{Shockley}. The magnetic Gr\"uneisen parameter divergence below 1~K~\cite{Singh} is possibly related to a small fraction moments which are not yet frozen. Another example is the site-diluted honeycomb iridate Li$_2$(Ir$_{0.45}$Ti$_{0.55}$)O$_3$ with magnetic frustration arising from a significant Kitaev interaction. Again a magnetic Gr\"uneisen parameter divergence is found despite the overall spin-glassy freezing of the moments and the non-divergence of the thermal Gr\"uneisen ratio~\cite{Manni}. 

Altogether these observations allow the following general conclusion. Thermodynamically, field sensitive quantum criticality requires a divergence of $\Gamma_H(T)$ at the critical field upon cooling. However, a divergence of $\Gamma_H$ towards $T,H \rightarrow 0$ may also be related to a small fraction of moments leading to a pronounced increase of $dM/dT$ upon cooling in low fields. To experimentally prove pressure sensitive zero-field quantum criticality the observation of a divergence of the thermal Gr\"uneisen ratio is therefore additionally required. In a couple of materials with divergent $\Gamma_H$ and respective scaling strong evidence against generic quantum criticality has been found. Either this points to pressure-independent ``strange metal'' behavior, or extrinsic origin due to magnetic disorder has to be considered.

\section*{Acknowledgments}

The results reviewed in this paper have been obtained in collaboration with Y. Tokiwa, C. Stingl, A. Sakai, S. Manni and Yogesh Singh. For collaboration and support with high-quality single crystals we acknowledge: E.D. Bauer (CeCoIn$_5$), M.-S. Kim and T. Takabatake (CeRhSn), J.J. Ishikawa and S. Nakatsuji (Pr$_2$Ir$_2$O$_7$), K. Kitagawa, K. Matsubayashi, M. Iwatani (YbCo$_2$Ge$_4$).

\end{document}